# Superconductivity in a new layered bismuth oxyselenide: LaO$_{0.5}$F0.5BiSe$_2$

A. Krzton-Maziopa[1], Z. Guguchia[2,3], E. Pomjakushina[4], V. Pomjakushin[5], R. Khasanov[3], H. Luetkens[3], P.K. Biswas[3], A. Amato[3], H. Keller[2] and K. Conder[4]

[1]*Faculty of Chemistry, Warsaw University of Technology, Noakowskiego 3, 00-664 Warsaw, Poland*
[2]*Physik-Institut der Universität Zürich, Winterthurerstrasse 190, CH-8057 Zürich, Switzerland*
[3]*Laboratory for Muon Spin Spectroscopy, Paul Scherrer Institute, 5232 Villigen PSI, Switzerland*
[4]*Laboratory for Developments and Methods, Paul Scherrer Institute, 5232Villigen PSI, Switzerland*
[5]*Laboratory for Neutron Scattering, Paul Scherrer Institute, 5232 Villigen PSI, Switzerland*

E-mail: anka@ch.pw.edu.pl

**Abstract**

We report superconductivity in a new layered bismuth oxyselenide LaO$_{0.5}$F$_{0.5}$BiSe$_2$ with the ZrCuSiAs-type structure composed of alternating superconducting BiSe$_2$ and blocking LaO layers. The superconducting transition temperature is T$_C$ = 2.6K, as revealed from DC magnetization, resistivity and muon spin rotation (μSR) experiments. DC magnetization measurements indicate a superconducting volume fraction of approximately 80%, which is at least twice higher in comparison to that found in corresponding sulphide LaO$_{0.5}$F$_{0.5}$BiS$_2$. Importantly, the bulk character of superconductivity in LaO$_{0.5}$F$_{0.5}$BiSe$_2$ was confirmed by μSR.

## I. Introduction

Recent discovery of superconductivity with T$_C$ around 8.6K in layered bismuth oxysulphide Bi$_4$O$_4$S$_3$ with ZrCuSiAs - structure type[1, 2] evoked intensive studies concerning the improvement of their superconducting properties. This material has a layered structure composed of superconducting (SC) BiS$_2$ layers and blocking layers of Bi$_4$O$_4$(SO$_4$)$_{1-x}$, where x indicates the defects of SO$^{2-}$ ions at the interlayer sites.

The sandwich structure of the SC and blocking layer is analogous to those of high temperature (high-$T_c$) cuprates and Fe-based superconductors. In both systems, $T_c$ can be enhanced by varying the blocking layers. The changing the blocking layer in $Bi_4O_4S_3$ resulted in a new $BiS_2$ based superconductor $LaO_{0.5}F_{0.5}BiS_2$. $LaO_{0.5}F_{0.5}BiS_2$ shows a small SC volume fraction for ambient pressure but achieves bulk superconductivity under high pressure with $T_c$ as high as 10.6 K. Those findings stimulated the scientific community for further investigations on different dopings and shortly after the first report on $LaO_{0.5}F_{0.5}BiS_2$ a set of new superconductors different substitutions with Ce, Pr, Nd, Yb for lanthanum appeared. [3, 4, 5, 6]

Superconducting $LnO_{1-x}F_xBiS_2$ materials are structurally related to the well-known layered iron based superconductor: $LaO_{1-x}F_xFeAs$[7], with the appearance of superconductivity after doping with fluorine. It has been found also that superconductivity in the $LaOBiS_2$ system can be induced not only by substituting oxygen for fluorine but also by increasing charge-carrier density (electron doping) through substitution of tetravalent elements, i.e. Th, Hf, Zr, and Ti for trivalent La[8]. By electron doping the parent phases of $LaOBiS_2$ and $ThOBiS_2$, considered as bad metals, become superconducting with $T_c$ of up to 2.85 K. On the other hand, an hole doping realized by substitution of divalent Sr for trivalent La did not induce superconductivity in those materials.

Another similarity of the $LnO_{0.5}F_{0.5}BiS_2$ compounds to the abovementioned layered pnictide and chalcogenide systems is the sensitivity of their SC properties to the change of structural parameters. The structure compression realized by an application of external pressure results in suppression of semiconducting behavior and enhances $T_c$ of $LaO_{0.5}F_{0.5}BiS_2$ to 10.6K, $PrO_{0.5}F_{0.5}BiS_2$ to 7.6K and 6.4K for $NdO_{0.5}F_{0.5}BiS_2$ in the same way as layered pnictides and chalcogenides do. With increasing pressure all the $BiS_2$ – type compounds show qualitatively similar evolution of superconducting transition temperature – an abrupt transformation from low $T_c$ phase to a high $T_c$ one with a characteristic maximum at pressures within the range from ca. 1.5 to about 2.5 GPa[9,10].

The existence of many common features related to crystal and band structure between BiS2 – based superconductors and pnictides/chalcogenides raised the question about the mechanisms of superconductivity in the newly discovered system. Theoretical considerations of density of states, band structure and Fermi surface nesting studied by means of the first principles calculations indicated that the insulating $LaOBiS_2$ parent phase (band gap of 0.15eV) transforms into metallic state after doping with fluorine[11]. At the optimal dopant content (x=0.5) there are four bands originating from the $p_x$ and $p_y$ orbitals of Bi, which cross the Fermi level. The surface nesting of $LaO_{1-x}F_xBiS_2$ was found to be weaker than that of corresponding pnictide LaOFeAs.

Superconductivity in the $BiS_2$-based compounds seems to be still under debate and even one of the reports indicates that superconducting response in $Bi_4O_4S_3$ is not a bulk phenomenon and might be impurity driven[12]. On the other hand recent publication about single crystals of $NdO_{1-x}F_xBiS_2$ grown by flux method[13] confirms clearly the bulk nature of superconductivity in BiS- type layered materials. An intensive work performed on different substitutions brought new $BiS_2$ – based materials with different substitutions for rare earth metals and in LnO – layer, however no attempts with substitutions in $BiS_2$ layer were tried up today. In the present work the new compounds $LaO_{0.5}F_{0.5}BiSe_2$, $LaO_{0.5}F_{0.5}BiTe_2$, and $LaO_{0.5}F_{0.5}SbS_2$ were synthesized and characterized using X-ray powder diffraction, neutron powder diffraction, resistivity, and magnetization experiments.

It was observed that replacement of sulfur by isovalent selenium results in a new superconductor LaO$_{0.5}$F$_{0.5}$BiSe$_2$. Superconducting properties of this compound was further investigated by means of μSR technique, which strongly indicates bulk superconductivity in LaO$_{0.5}$F$_{0.5}$BiSe$_2$. The compounds LaO$_{0.5}$F$_{0.5}$BiTe$_2$ and LaO$_{0.5}$F$_{0.5}$SbS$_2$ do not reveal superconductivity down to T=1.7K.

**II. Experimental**

Polycrystalline samples of LaO$_{0.5}$F$_{0.5}$BiSe$_2$ were obtained via solid state reaction from high purity (at least 99.99%, Alfa) powders of La$_2$O$_3$, LaF$_3$ and Se. Appropriate amounts of lanthanum metal and bismuth pieces were used to keep the proper molar ratio of reacting species. Starting materials with nominal composition LaO$_{0.5}$F$_{0.5}$BiSe$_2$ were weighted in the He-filled glove box, carefully mixed, pressed into pellets, sealed in evacuated quartz ampoules and preheated at 800ºC for 15 hours. Afterwards the products were transferred to glove-box, thoroughly ground in the mortar under helium atmosphere, pressed again into pellets and re-sealed under vacuum in quartz ampoules. Next the ampoules were heated again at 800ºC for 48h followed by cooling to 300ºC and quenching in air. To avoid any contamination form environment the as synthesized materials were transferred to glove-box and kept in inert atmosphere.

The phase purity of the as prepared materials was characterized by powder X-ray diffraction (XRD) using a D8 Advance Bruker AXS diffractometer with Cu K*α* radiation. For these measurements a low background airtight specimen holder was used. The samples were additionally studied by means of neutron powder diffraction (NPD) at the SINQ spallation source of the Paul Scherrer Institute (PSI, Switzerland) using the high-resolution diffractometer for thermal neutrons, HRPT,[14] with the neutron wavelengths λ = 1.494 and 1.886 Å. To avoid degradation in air the samples were loaded into a vanadium containers with an indium seal in an He glove box. The Rietveld refinements of the crystal structure parameters were done using the FullProf package[15] with the use of its internal tables.

The magnetic susceptibility was measured by a SQUID magnetometer (*Quantum Design* MPMS-XL). Temperature dependent measurements of the resistivity were carried out using a standard four-probe method in a Physical Property Measurement System (*Quantum Design* PPMS).

Transverse-field (TF) $\mu$SR experiments were performed at the $\pi$M3 beamline of the Paul Scherrer Institute (Villigen, Switzerland), using the general purpose instrument (GPS). The sample was mounted inside of a gas-flow $^4$He cryostat on a sample holder with a standard veto setup providing essentially a low-background $\mu$SR signal.

**III. Results and discussion**

**A. Crystal structure and sample quality**

The samples after sintering are black in color and hard in nature. X-ray powder diffraction and neutron powder diffraction studies of the as sintered material revealed that except a main phase $LaO_{0.5}F_{0.5}BiSe_2$ with tetragonal ZrCuSiAs –type structure fitted to P4/nmm crystal metric with $a$ = 4.15941(7) and $c$ = 14.01567(34) Å. The as grown samples contain also impurity phases such as traces of unreacted starting materials i.e. $La_2O_3$ and Bi, and possibly $Bi_2Se_3$ or other binary Bi-Se compounds. Layered structure of the new compound resembles similar features as those of sulphur analogue and consists of alternating $La_2(OF)_2$ (fluorine doped rare earth oxide layer) and $BiSe_2$ (fluorite type) sheets. Atomic coordinates, Wyckoff positions and site occupancies for studied samples are listed in Table1.

**Table 1.** Atomic coordinates, Wyckoff positions, atomic displacement parameters B, and site occupancy of $LaO_{0.5}F_{0.5}BiSe_2$. Space group P4/nmm

| $LaO_{0.5}F_{0.5}BiSe_2$ | x | y | z | B, (Å$^2$) | site | occupancy |
|---|---|---|---|---|---|---|
| La | 0.2500 | 0.2500 | 0.0943(4) | 1.3(1) | 2c | 1 |
| O  | 0.7500 | 0.2500 | 0.0000(0) | 1.3(1) | 2a | 0.5 |
| F  | 0.7500 | 0.2500 | 0.0000(0) | 1.3(1) | 2a | 0.5 |
| Bi | 0.2500 | 0.2500 | 0.6206(4) | 0.8(1) | 2c | 1 |
| Se1 | 0.2500 | 0.2500 | 0.3847(4) | 1.2(9) | 2c | 1 |
| Se2 | 0.2500 | 0.2500 | 0.8115(3) | 1.2(9) | 2c | 1 |

## B. Electrical resistivity

Fig. 1a shows the temperature dependence of resistivity for $LaO_{0.5}F_{0.5}BiSe_2$ in the temperature range between 1.8 K and 300 K.

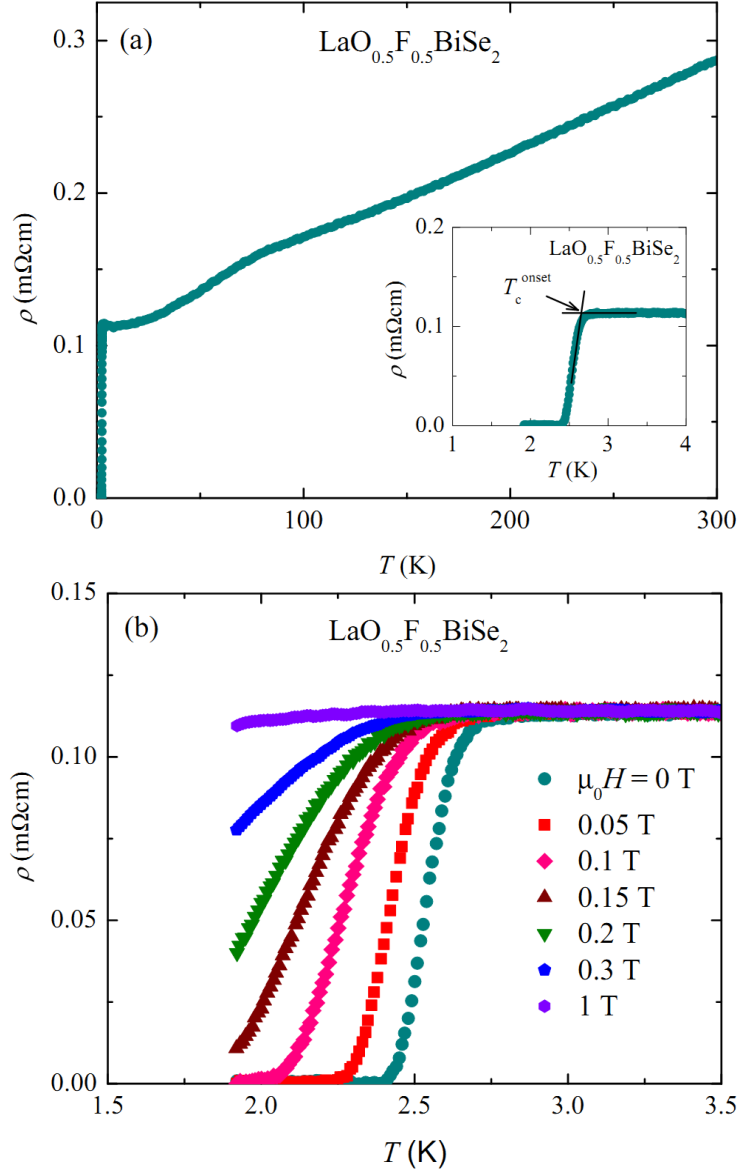

**FIG. 1.** (Color online) (a) Temperature dependence of the electrical resistivity $\rho$ of polycrystalline $LaO_{0.5}F_{0.5}BiSe_2$ sample. The inset shows the data near $T_c$ at low temperatures. (b) Low temperature resistivity data under magnetic fields up to 1 T.

The metallic behavior is observed above 5 K state with the residual resistivity value of 0.12 mΩcm. The transition to the superconducting state is clearly visible below $T_c^{onset} \approx 2.7$ K, as illustrated in the inset of Fig. 1a. Note that the metallic behavior of the normal state was also observed in the superconductor $Bi_4O_4S_3$. However, for this compound much higher value for the residual resistivity 2 mΩcm was reported. The metallic character of resistivity in $LaO_{0.5}F_{0.5}BiSe_2$ is in contrast to the semiconducting behavior observed for corresponding $LaO_{0.5}F_{0.5}BiS_2$, widely described in the literature. Note that the new compounds $LaO_{0.5}F_{0.5}BiTe_2$ and $LaO_{0.5}F_{0.5}SbS_2$ were also synthesized, but no superconductivity was observed down to 1.7 K. Low temperature resistivity data recorded in zero field and under applied magnetic fields up to 1 T for $LaO_{0.5}F_{0.5}BiSe_2$ are presented in Fig.1b. The superconducting states are destroyed by applying high magnetic fields.

**C. Magnetization**

To determine whether superconductivity is a bulk phenomenon in $LaO_{0.5}F_{0.5}BiSe_2$, zero field-cooled (ZFC) and field-cooled (FC) magnetic susceptibility $\chi$ was measured in a magnetic field of $\mu_0 H = 0.5$ mT. The results shown in Fig.2 evidence sharp superconducting onset at 2.6 K as indicated by a vertical gray line. The value of ZFC susceptibility at 1.8 K corresponds to about 80 % volume fraction of superconductivity. It is clear that, even though the superconducting transition is incomplete at 1.8 K the volume fraction at that temperature is appreciable. The strong diamagnetic signal below $T_c$ is consistent with bulk superconductivity in $LaO_{0.5}F_{0.5}BiSe_2$.

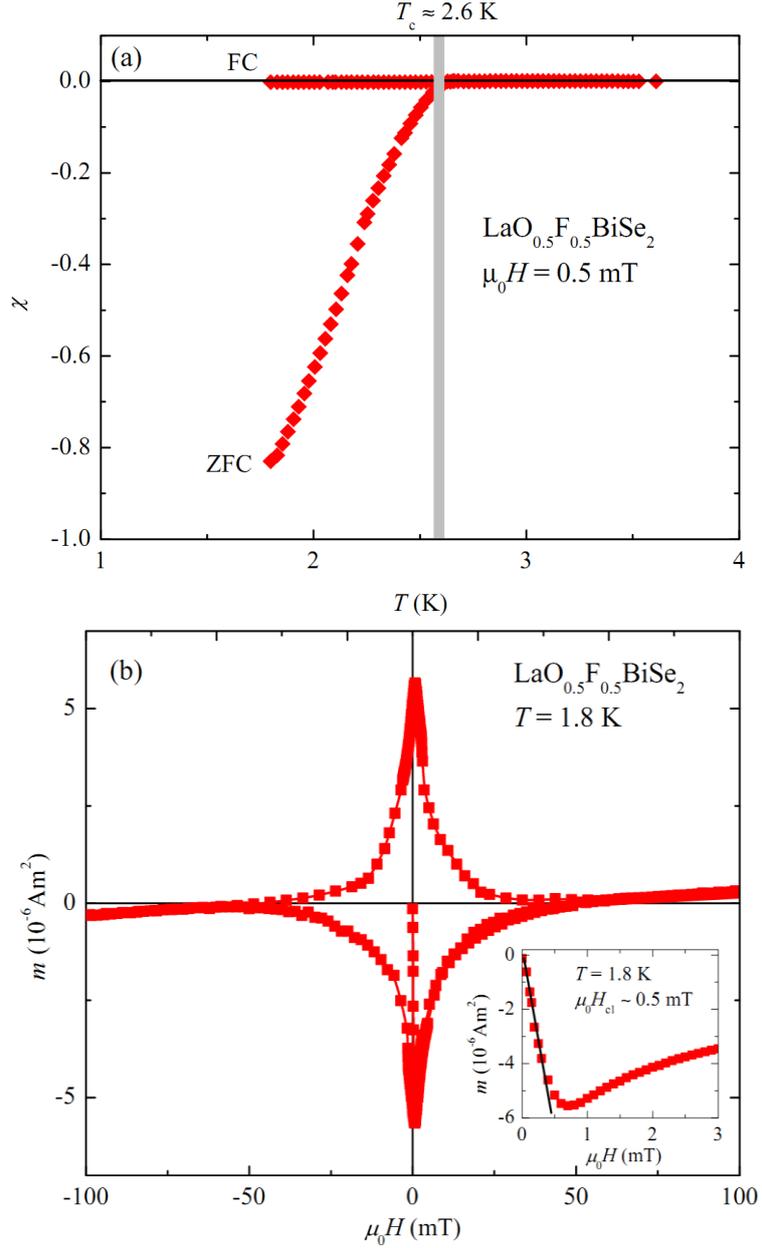

**FIG. 2.** (Color online) (a) Temperature dependence of the ZFC and FC magnetic susceptibility of $LaO_{0.5}F_{0.5}BiSe_2$ in a magnetic field of $\mu_0 H = 0.5$ mT. The vertical gray line denotes the superconducting transition temperature $T_c$. (b) Field dependence of the magnetic moment at $T = 1.8$ K in an applied field up to $\mu_0 H = 100$ mT. The inset shows the low field data.

The field dependence of the magnetic moment of $LaO_{0.5}F_{0.5}BiSe_2$ was studied at the base temperature $T = 1.8$ K for external magnetic fields up to $\mu_0 H = 100$ mT and shown in Fig. 2b. The inset demonstrates that the initial flux penetration and the deviation from linearity determine the lower critical field of this compound $\mu_0 H_{c1} =$ 0.5 mT. The second critical field was found to be also small: $\mu_0H_{c2}$ = 60 mT. Note, that wide open magnetic moment hysteresis $m(H)$ loop of the studied compound demonstrates its bulk superconductivity.

### D. Muon spin rotation measurements

Fig. 3 exhibits the transverse-field (TF) muon-time spectra for $LaO_{0.5}F_{0.5}BiSe_2$ measured in an applied magnetic field of $\mu_0H$ = 11.5 mT in the SC state at 1.5 K.

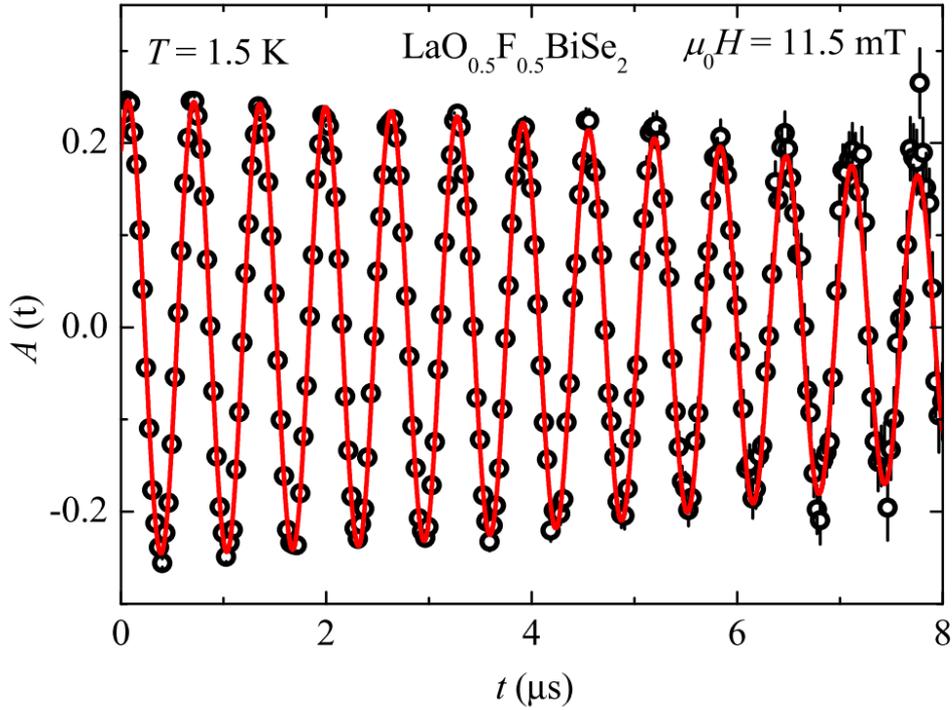

**FIG. 3.** (Color online) Transverse-field (TF) $\mu$SR time spectrum obtained in $\mu_0H$ = 11.5 mT in the SC state at $T$ = 1.5 K. The solid line represents the fit to the data by means of Eq. 1.

The data were analyzed by using the following functional form:

$$P(t) = A\exp\left[-\frac{(\sigma^2_{sc}+\sigma^2_{nm})t^2}{2}\right]\cos(\mu_0\gamma_\mu H_{int}t+\varphi), \qquad (1)$$

Here $A$ denotes the initial asymmetry, $\gamma/(2\pi) \approx$ 135.5 MHz/T is the muon gyromagnetic ratio, and phi is the initial phase of the muon-spin ensemble. $\mu_0H_{int}$ represents the internal magnetic field at the muon site, and the relaxation rates $\sigma_{sc}$ and $\sigma_{nm}$ characterize the damping due to the formation of the FLL in the superconducting state and of the nuclear magnetic dipolar contribution, respectively. In the analysis $\sigma_{nm}$ was assumed to be constant over the entire temperature range and was fixed to the value $\sigma_{nm} = 0.114(2)$ µs$^{-1}$ obtained above $T_c$ where only nuclear magnetic moments contribute to the muon depolarization rate σ. As indicated by the solid lines in Fig. 4, the µSR data are well described by Eq. (1). The temperature dependence of $\sigma_{sc}$ for LaO$_{0.5}$F$_{0.5}$BiSe$_2$ is shown in Fig. 4a. Below $T$c ≈ 2.6 K, the relaxation rate $\sigma_{sc}$ starts to increase from zero due to the presence of a nonuniform local field distribution as a result of the formation of a flux-line lattice (FLL) in the SC state. It is worth mentioning that the value of σ in the SC state is rather small, which implies that the magnetic penetration depth λ, which is one of the fundamental parameters of a superconductor, is sufficiently large (since the data points for σ below 1.5 K are missing in the present work, it is not possible to determine the zero temperature limit of σ, which would allow to determine λ(T=0)). Note that recent detailed µSR studies[16] on the similar system Bi$_4$O$_4$S$_3$ revealed that it also exhibits one of the highest λ among all other superconductors.

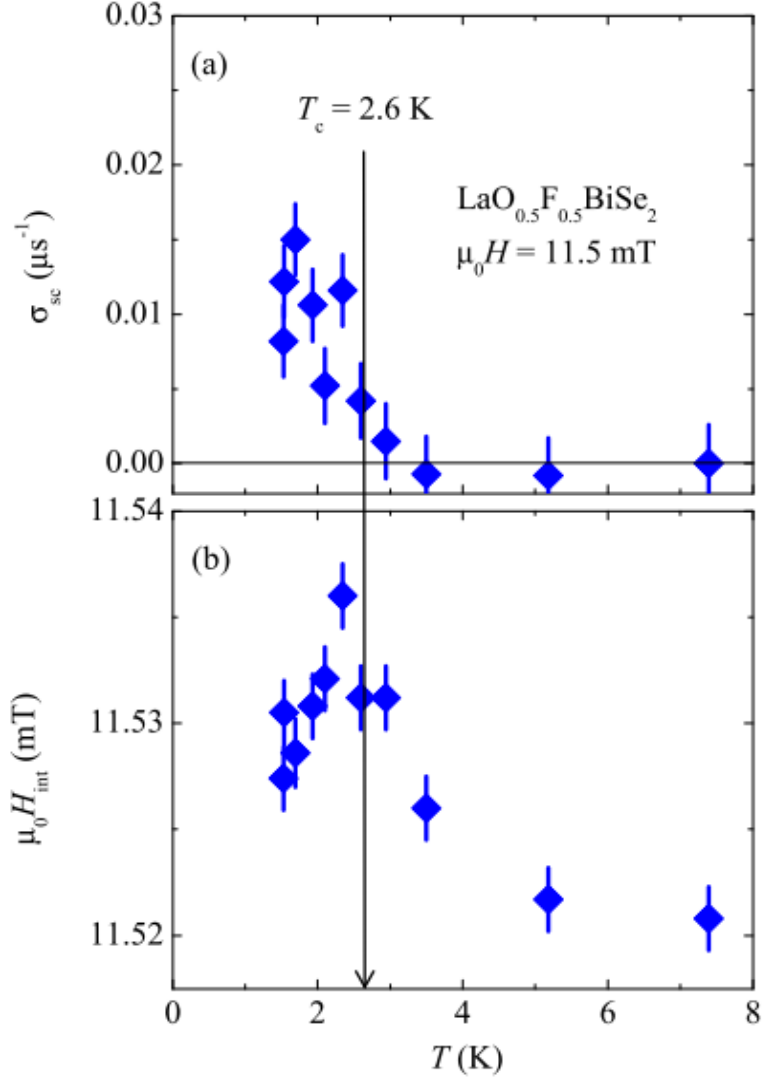

**FIG. 4.** (Color online) Temperature dependence of the superconducting muon spin depolarization rate $\sigma_{sc}$ (a) and internal field (b) measured in an applied magnetic field of $\mu_0 H = 11.5$ mT for $LaO_{0.5}F_{0.5}BiSe_2$. An arrow denotes the superconducting transition temperature $T_c$.

In Fig.4b the temperature dependence of the $\mu_0 H_{int}$ is presented. $\mu_0 H_{int}$ increases with decreasing temperature in the normal state, which may be caused by an enhancement of the paramagnetic susceptibility of the sample $LaO_{0.5}F_{0.5}BiSe_2$ upon lowering the temperature. Decrease of the internal field $\mu_0 H_{int}$ (diamagnetic shift) was observed below $T_c \approx 2.6$ K, which is evident from Fig. 4(b). Observed diamagnetic shift as well as the increase of $\sigma$ confirms the bulk character of superconductivity in $LaO_{0.5}F_{0.5}BiSe_2$.

## 4. Summary

In conclusion we report on a synthesis, crystal structure and SC properties of a new Se-containing layered superconductor: $LaO_{0.5}F_{0.5}BiSe_2$, obtained by replacement of sulphur for selenium in recently reported $BiS_2$ – type superconductors. By this substitution the volume of the crystal unit cell increases and the c-parameter expands from 13.3157 Å to 14.01567Å. The new superconductor exhibits bulk superconductivity with $T_c$ = 2.6K. DC magnetization measurements revealed a superconducting shielding fraction of approximately 80 %, which is at least twice higher in comparison to that found in corresponding sulphide $LaO_{0.5}F_{0.5}BiS_2$. Moreover, bulk superconductivity in $LaO_{0.5}F_{0.5}BiSe_2$ was confirmed by μSR experiments.

No superconducting transition was observed in analogous phases: $LaO_{0.5}F_{0.5}SbS_2$ with antimony substituted for bismuth and $LaO_{0.5}F_{0.5}BiTe_2$ with tellurium for Se.

## Acknowledgements

This work has been supported by the European Union in the framework of European Social Fund through the Warsaw University of Technology Development Program.